# CDW and similarity of the Mott-Insulator-to-Metal transition in cuprates with the gas to liquid-liquid transition in supercooled water

G. Campi[1,2], D. Innocenti[3], A. Bianconi[1,2]

[1]*Institute of Crystallography, CNR, Via Salaria Km 29.300, 00015 Monterotondo Rome, Italy*
[3]*EPFL Laboratory of Nanostructures and Novel Electronic Materials, SB ICMP LPMC PH D2 475 (Bâtiment PH) Station 3, CH-1015 Lausanne Switzerland*
[2]*Rome International Centre for Materials Science Superstripes RICMASS, via dei Sabelli 119A, 00185 Roma Italy*

**Abstract**

New advances in x-ray diffraction, extended x-ray absorption fine structure EXAFS and x-ray absorption near edge structure XANES using synchrotron radiation have now provided compelling evidence for a short range charge density wave phase (CDW) called "*striped phase*" in the $CuO_2$ plane of all cuprate high temperature superconductors. The CDW is associated with a bond order wave (BOW) and an orbital density wave (ODW) forming nanoscale puddles which coexist with superconducting puddles below Tc. The electronic CDW crystalline phase occurs around the hole doping 0.125 between the Mott charge transfer insulator, and the 2D metal. The Van der Waals (VdW) theoretical model for a liquid of anisotropic extended objects proposed for supercooled water is used to describe : a) the underdoped regime as a first spinodal regime of a "slightly doped charge transfer Mott insulator puddles coexisting with the striped polaronic CDW puddles; and b) the optimum doping regime as a second spinodal regime where striped polaronic CDW puddles coexist with the normal 2D metal puddles. This complex phase separation with 3 competing phases depends on the strength of the anisotropic electron-phonon interaction that favors the formation striped polaronic CDW phase.

**Key works**: High Temperature Superconductivity - Phase separation -CDW - Charge Density Wave

1. **Introduction**

Transition metal oxides are commonly termed as 'complex oxides' since they give rise to a variety of phenomena, various charge, spin, orbital and lattice fluctuations giving ferroelectrics, colossal magneto resistence materials, relaxors, and multiferroics. They have important applications as dielectrics, semiconductors as materials for magnetic and optical applications. The discovery of high temperature superconductivity (HTS) has highlighted that a novel quantum phase of condensed matter appears in complex transition metal oxides. This rich phenomenology arises from the complex structural properties of these systems, well described in their phase diagrams. Intrinsic inhomogeneity and complexity in high temperature superconductivity appears nowadays to be a key point to unveil this quantum phase of matter.

The electron crystalline phase called Charge Density Wave (CDW) is well known to occur in low-dimensional organic conductors and layered systems [1-5]. The charge density wave (CDW) is accompanied by a periodic lattice distortion (PLD) i.e., a bond distance modulation (BOW) and an orbital density wave (ODW) in the presence of a sizeable electron-phonon coupling as in a Peierls transition. The CDW can be due to nesting of the Fermi surface or to long range coulomb repulsion stabilizing a Wigner lattice of localized charges. A polaronic CDW competing with a Luttinger liquid appears in a strongly correlated low density electronic liquid in presence of electron-phonon interaction [6-7].

The first experimental evidence of a short range CDW in cuprates superconductors was provided in 1991 by extended x-ray absorption fine structure (EXAFS) and X-ray absorption near edge structure (XANES) joint with detetection of diffuse short range reflections in x-ray diffraction using synctrotron radiation [8-34]. EXAFS and XANES [35-37] provide direct evidence for the amplitude of periodic lattice distortion (PLD) beeing fast probes sensitive to short range lattice structural order. Because of these features EXAFS and XANES [39-40] have been succesfully used to probe short range CDW in a variety of compounds [42-56]. Tuning the photon energy at the Cu K-edge [38,39] and at the Cu $L_3$ edge [41] of cuprates it has been possibile to detect resonant X-ray diffraction (XRD) due to displacenents of selected atomic species obtaining direct evidence of CDW in cuprates [57-61]. Recently weak diffuse reflections due to CDW have been revealed also by XRD using detectors of high sensitivity and a focused micro X-ray beam [62-69] in nearly all cuprate families. The polaronic character of the CDW has been shown by the giant isotope effect on its temperature onset [70-72] and by the role of the lattice misfit strain between the active copper oxide layer and the spacer layers [73-77].

The characteristic feature of the CDW phase in cuprates is a nanoscale phase separation where below T* nanoscale puddles of CDW coexists with magnetic puddles and below Tc the short range CDW phase competes and coexists with the high temperature superconducting phase. This well accepted complex nanoscale phase separation [78-79] has been called the superstripes scenario [80-82] and it has been observed also in doped iron based superconductors [83] and diborides [84]. The coexistence of two different electronic components, phase separation, lattice and electronic complexity and percolation have been predicted by several authors to be essential features for emergente of high tenperature superconductivity [85-99].

The superstripes phase is a case of arrested phase separation made of a type of nanoscale phase separation, forming complex multiscale pattern from atomic scale to nanoscale (1-100 nm) to mesoscale (100-1000 nm) to microscale (1-100 microns). These nanoscale ripples puddles or striped puddles are associated with orbital angular momentum modulation, modulated local lattice distortions, charge modulation, spin modulation, showing a chiral order [100-103] The transition from the high temperature disordered phase to the ordered phase of SDW, ODW, CDW is very broad due to the ineluctable presence of defects like oxygen interstitials, impurities, and lattice distortions due to the misfit strain between the copper oxide layer and the spacer layers [104-111] which control the arrested phase transition in cuprates as weel as in cobaltites [112]. The complex networks of defects have a heterogeneous degree distribution following a power law with a finte cut off. The presence of a fat tail reveals heterogeneity which is a signal of self organization with the presence of hubs. The deviation from randomness may be attributed to self organization and when the structure deviates fron randomness has strong effects on the dynamics that take place on top of networks.

There is now agreement that this striped CDW phase occurs at doping δ=1/8 as an intermediate phase between the Mott insulator, at δ=0, and the normal metal at δ=1/4. In this doping range, a first well established spinodal phase separation in cuprates occurs in the underdoped regime, 0<δ<1/8, where the slightly doped Mott insulator coexists with striped *CDW puddles* below $T^*$. A second spinodal phase separation occurs at optimum doping, 2/16<δ<4/16, showing a striped CDW phase coexisting with the *normal metallic* puddles. This *normal metallic matter* is the only component present at doping higher than 0.25. Therefore the high $T_c$ superconducting phase appears in two different regimes of phase coexistence: at a low doping regime, 1/16<δ<2/16, where both the Mott insulator and the striped matter coexist, and at higher doping regime, 2/16<δ<4/16, where there is coexistence of CDW phase and the metallic phase.

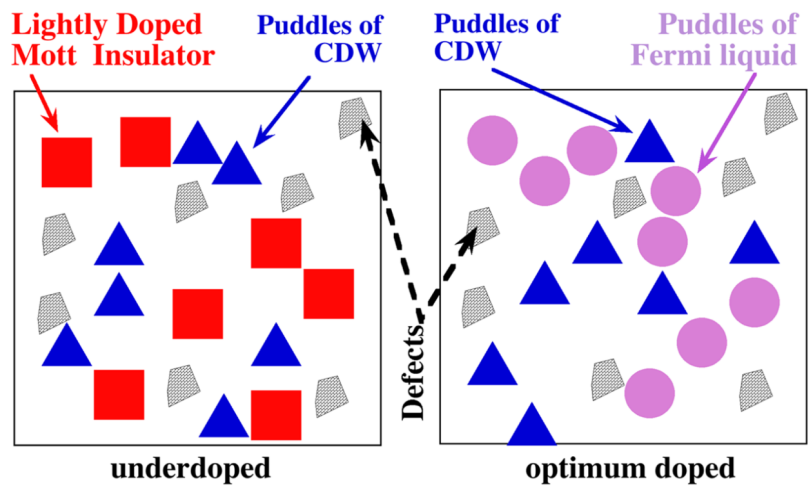

**Fig.** 1 Pictorial view of nanoscale phase separation in the underdoped regime for doping δ<1/8 and in the optimum doping regime for doping δ>1/8 holes per Cu site.

This latter can be further divided into two different regimes: the optimum doping regime 2/16<δ<3/16 where the striped cdw matter is dominant, and the overdoped regime 3/16<δ<4/16 where the normal metal is dominat. The optimum doping regime for the highest $T_c$ around 3/16 is cleary a percolation regime as shown in Figure 1.

## 2. The generalized Van der Waals thermodynamic model

In order to include the CDW phase at 1/8 doping in the generic phase diagram of the doped Mott insulator we have extended the Van der Waals (VdW) model proposed by Emin [113]. It is weel established that the undoped parent compounds are strongly correlated Mott insulators with a Hubbard charge transfer gap. It is now well established that the Mott to insulator transion occurs by inserting the doped charge carriers (characterized by the many body configuration $3d^9\underline{L}$) in the Hubbard gap which remains nearly constant in the crossover region[82] for doping less than ¼ holes per Cu site. At very low doping the itinerant carriers are $3d^9\underline{L}$ states can be considered as a low density gas. Because of a the sizeable electron-phonon interaction the doped charges form a low large polarons [26,6,7,71,72,75,76,86,87] or bipolarons [99,113,114] of radius R. This polaronic gas in the lightly doped cuprate is characterized by a short range repulsion between polarons and a long range phonon mediated attraction[113]. Therefore increasing the polaron density a gas to liquid transition similar to Van der Walls gas to liquid transition in a real gas is expected. In the Wan der Walls model the b and a constants respectively characterize the short range repulsion and the long range attraction. However the polaronic charge is an extened object with a material dependent anisotropic interaction, due to the mixing between different orbital moments[13-19], which promotes the formations of polaronic stripes and a polaronic charge density wave[12-35,86,] at doping δ=1/8 smaller than the density of the homogeneous Fermi liquid forming at δ=2/8.

For isotropic polarons the curve of coexistence of the very low density gas (VLD) δ≈0 with liquid, δ≈1/4 is obtained by the Maxwell construction of an effective VdW state equation for polaronic charges. In that model the parameter "*b*" of the VdW equation is the volume occupied by a polaron and "*a*" is the effective polaron-polaron attractive interaction. Starting from this view, the CDW phase is introduced by a new term taking into account an electron-phonon anisotropy interaction. This new parameter is the directional interaction between particles and favours the formation of short range unidimensional CDW. Therefore, in order to describe the phase diagram of cuprate superconductors we include an additional term to the standard VdW scheme, analogous to that introduced for supercooled water [115] by Poole at al. [116]. This term consists of a characteristic energy "$E_\gamma$" for the formation of *striped CDW* at a preferential volume $V_\gamma$ =1/8 [117]. In supercooled water the introduction of these new term provides a complex phase diagram with the occurrence of phase separation between a high density liquid (HDL) (which corresponds to cuprate

Fermi liquid) and a low density liquid (LDL) (which corresponds to the striped CDW phase) when $E_\gamma$ becomes larger than the attraction, $E_{VdW}$. Thus, the tendency to form striped CDW of doped holes is described by the free energy term $A_\gamma$, added to the VdW free energy $A_{VDW}$ yielding the total free energy:

$$A = A_{VDW} + A_\gamma \quad (1)$$

where $A_{VDW}$ and $A_\gamma$ are given by

$$A_{VDW} = -RT\left\{\ln\left[(V-b)/\Lambda^3\right]+1\right\} - a^2/V \quad (2)$$

$$A_\gamma = -fRT\ln[\Omega + \exp(-E_\gamma/RT)] - (1-f)RT\ln(\Omega+1) \quad (3)$$

In the last expression $f$ describes the range of optimum density for the striped CDW matter given by

$$f = \exp\left\{[(V-V_\gamma)/\sigma]^2\right\} \quad (4)$$

Here the constant of the VdW model corresponding to a molecular volume, $b$, is taken to be the volume filled by the dressed polaronic charge carriers. The VdW force constant, associated with the inter-molecular attraction, $a$, provides a central interaction over the characteristic constant volume, with energy $E_{vdw}=a/b$, while $E_\gamma$ is defined as the energy due to the anisotropic interactions, generating directional bonds. In this model there are $\Omega \gg 1$ configurations all having $E_\gamma=0$ and only a single configuration in which is allowed the formation of directional bonds with energy $E_\gamma$. These directional bonds are most likely to occur when the bulk molar volume has the value $V_\gamma$. This is consistent with each charge carrier having the optimal local volume for the formation of directional bonds to its neighbours [117,118].

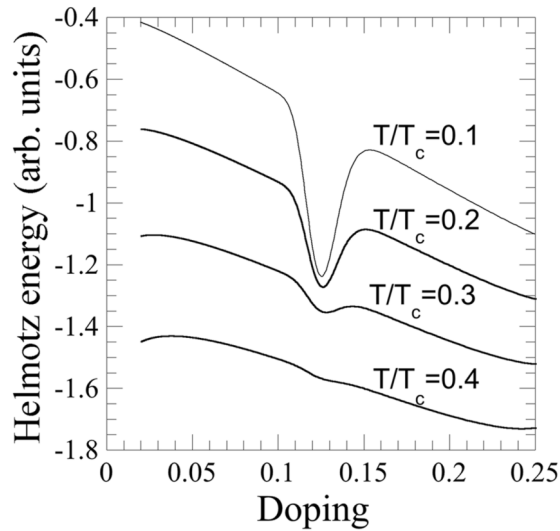

**Fig. 2** The Helmotz energy as a function of doping, δ, for different temperatures of $T/T_c$, where $T_c=8a/27Rb$ is defined as in the Van der Waals scheme.

However, when $V \neq V_\gamma$, the directional bonds are only a fraction $f$ of the total, since $V$ is no longer consistent with the possibility that all directional bonds occur in the optimal local volume. The remaining fraction $1-f$ of bonds occurs in unfavourable local volume and thus do not have the potential to form directional bonds. The parameter $\sigma$ describes the width of the region of $V$ around $V_\gamma$ over which a significant fraction of directional bonds follows the Eq. (1).

The strength of the anisotropic interaction is temperature dependent, as we show in Fig. 2, where the minimum of the Helmotz free energy at $\delta \sim 1/8$ becomes deeper with decreasing temperature. That indicates a favored formation of the intermediate phase **[2]** at low temperatures.

The inclusion in the model of an optimum volume for directional bonding produces a phase separation with the formation of an intermediate CDW phase **[2]** when $V$ tends to $V_\gamma$. The stability of the CDW intermediate phase depends on the energy of the anisotropic interaction, $E_\gamma$, in comparison with the classical CDW isotropic interaction $E_{vdw}$. Indeed, in the case $E_\gamma > E_{vdw}$, the anisotropic interactions prevails and the system favours the formation of directional bonds. As a consequence, the new intermediate CDW phase **[2]**, becomes stable and the homogeneous *striped CDW matter* occurs at $\delta \sim 1/8$. The normal VdW spinodal curve splits into two spinodal lines **[1+2]** and **[2+3]**, each terminating at a critical point shown in Fig. 3.

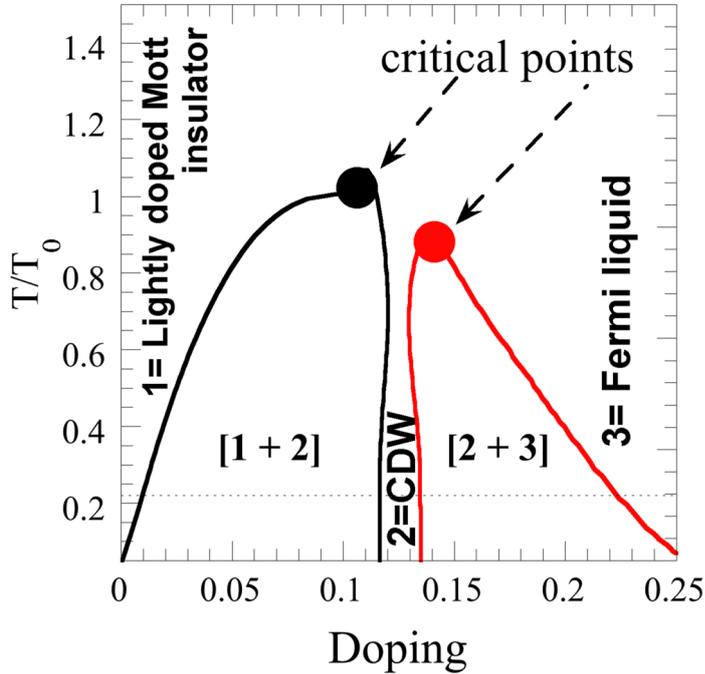

**Fig. 3** The phase diagrams obtained by computing the spinodal lines from Eq. (1) for $E_\gamma > E_{vdw}$. We can observe the occurrence of two phase separations indicated by the two spinodal lines.

The first spinodal region "underdoped regime" occurs for δ smaller than 1/8 where the partial density of first component **[1]**, the antiferromagnetic slightly doped Mott insulator coexists with the intermediate CDW striped phase **[2]**.

The insulator-to-metal phase transition in cuprates at doping 1/16 is assigned to the percolation threshold for macroscopic networls of striped CDW puddles. In the second spinodal region for doping larger than 1/8, the puddles of normal metal with the conventional Fermi liquid **[3]**, coexists with the striped CDW phase **[2]**. The "optimum doping" regime corresponds to the range where the striped phase and the normal metal, partial densities, are equal as shown in figure 4. The range called "overdoped" regime is where the partial density of the normal metal becomes larger than the partial density of the striped liquid. Different probability distributions $\rho_{[1]}$, $\rho_{[2]}$ and $\rho_{[3]}$ of hole sites in three different phases **[1]**, **[2]** and **[3]** are illustrated in Fig. 4.

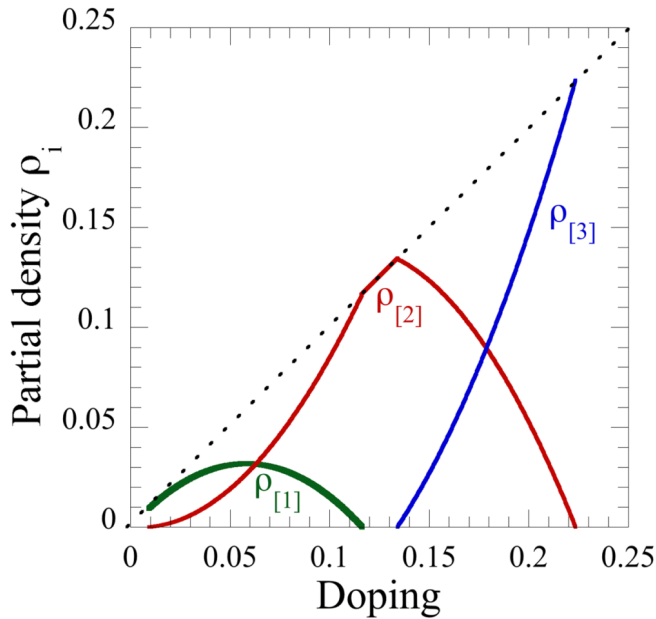

**Fig. 4** The calculated partition of the number density of the doped holes in the partial number density of the first component $\rho_{[1]}$, the slightly doped Mott insulator, of the second component $\rho_{[2]}$, the striped CDW matter, and of the third normal Fermi liquid , $\rho_{[3]}$, as a function of doping at the fixed temperature of T=18.24 K.

This scenario was first observed in cuprate families as $La_{1.6-x}Nd_{0.4}Sr_xCuO_4$ (LNCO214) where CDW with an associated SDW phase appears at δ=1/8, for temperatures lower than the pseudogap temperature T*(δ). In order to fit the experimental data of the pseudogap temperature T*(δ) by using our model, we extrapolate the parameter *b* from the value of the minimum density (*1/b*=0.28 holes per Cu site) for the homogeneous non-superconducting normal phase, where T* (δ) goes to zero. Moreover we impose $V_\gamma/b$=2.24 and $\sigma=V_\gamma/10$ to obtain the appearance of the intermediate phase at δ=1/8 , and $\Omega=exp(-S_\gamma/R)$ where $S_\gamma$ =-70.25 J/(K mol) is the formation entropy of a mole of directional bonds. We show that the phase diagram of the normal phase of cuprate below T* can be obtained by setting properly the value of $E_\gamma/E_{vdw}$ ratio.

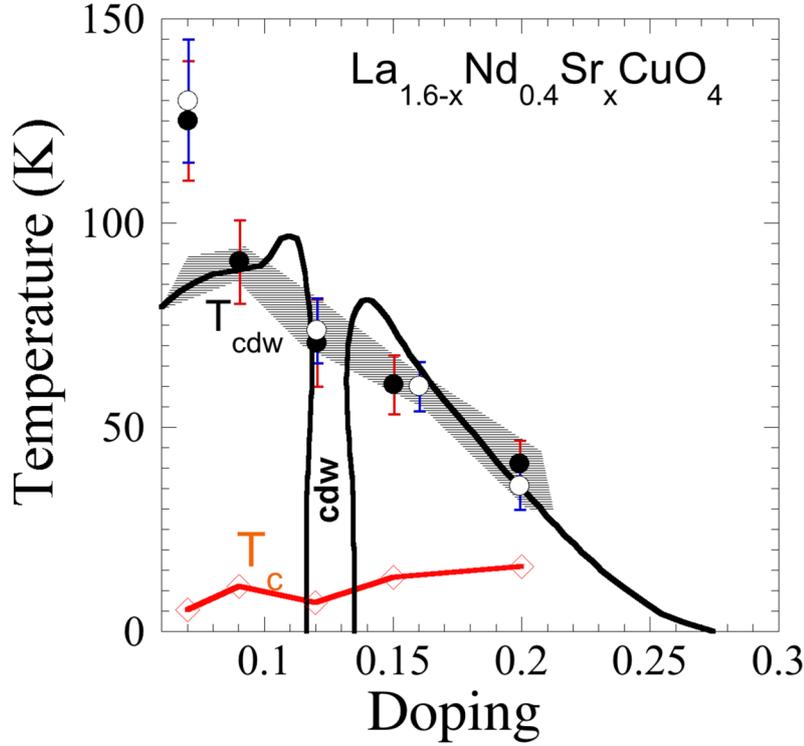

**Fig. 5** The phase diagram for case of $La_{1.6-x}Nd_{0.4}Sr_xCuO_4$. We have used $E_{vdw} = (1.68 \pm 0.17)$ KJ/mol and $E_\gamma = 2.2$ KJ/mol ($E_\gamma/E_{vdw} \sim 1.3$). The occurrence of the CDW phase at $\delta=1/8$ is illustrated. The characteristic T* temperatures measured by $^{63}Cu$ NQR experiments and XANES experiments are indicated by the full and open circles respectively. The open diamonds correspond to the superconducting $T_c$.

In Fig. 5 we report the phase diagram of LNCO214 resulting from the calculation of the spinodal line from Eq. 1. We have imposed $E_{vdw}=(1.68\pm0.17)$ KJ/mol and $E_\gamma=(2.2\pm0.2)$ KJ/mol ($E_\gamma/E_{vdw}\sim1.3$) in order to fit the T* temperatures measured by XANES and $^{63}Cu$ NQR experiments for $La_{1.6-x}Nd_{0.4}Sr_xCuO_4$ (full circles) and $La_{1.8-x}Eu_{0.2}Sr_xCuO_4$ (open circles). Here, being $E_\gamma>E_{vdw}$, the effect of the $A_\gamma$ term in Eq. 1, related to the strength of the directional bonds, is to "split" the normal liquid-gas spinodal curve by imposing thermodynamic stability in the region of states centered on $\delta=1/8$. As a result, two spinodal lines occur, each one terminating at two critical points, denoted $T^*_{C(1)}$ and $T^*_{C(2)}$ shown in Fig. 3. The CDW phase appearing at $\delta\sim1/8$ contributes to two spinodal lines : the first spinodal line, ($T<T^*_{C(1)}$), where the magnetic slightly doped Mott insulator coexists with the CDW phase. In the second spinodal line below $T^*_{C(2)}$ ($T<T^*_{C(2)}$) the striped CDW puddles coexists with the high density normal metal puddles.

## 3. Conclusions

We have presented a simple thermodynamic model for the phase diagram of the electronic complex liquid where two spinodal coexistence regimes are present. In the first one, an antiferromagnetic spin fluid coexists with a striped liquid of quasi linear chains network and in the second that *striped* liquid coexists with a normal liquid `supporting our earlier proposal [117]`.. To describe the coexistence of nanoscale striped *pseudogap matter* and normal matter in cuprate compounds we have used the Poole-Sciortino approach. In particular it gives account of the following features in cuprates:

(1) the stability around $\delta \sim 1/8$ of the CDW striped phase

(2) the coexistence below T* in the underdoped regime of a first phase made of first puddles of slightly doped Mott insulator and second puddles of striped CDW striped puddles

(3) the coexistence below T* in the optimum doping regime of first phase made of first puddles of striped CDW striped puddles and second puddles of normal Fermi liquid.

(4) qualitative reproduction of the curves of T*($\delta$) in a cuprate family.

This work provides a possible phase diagram of a multi-component scenario. The simplicity of this approach makes the model useful for understanding the phase diagram of all cuprates. Furthermore, The complex phase separation with two different spinodal regimes, indicates that the physics of high $T_c$ superconductors resembles the new physics of complex systems like biological matter characterized by multiscale structure and dynamical fluctuations between different local conformations. In this scenario the mechanism of high temperature superconductivity in cuprates clearly requires a theory including multiple electronic components, multi condensates and quantum confinement [118-120].